\documentclass[12pt,preprint]{aastex}
\usepackage[onecolumn,numberedappendix]{emulateapj5}
\usepackage{epsfig}
\usepackage{rotating}      
\tightenlines
\slugcomment{Submitted to the Astrophysical Journal}

\newcommand     \beq    {\begin{equation}}

\newcommand     \eeq    {\end{equation}}

\newlength{\figwidth}
\countdef\decade=200
\advance\decade by \year
\countdef\hours=201
\advance\hours by \time
\divide\hours by 60
\countdef\mins=202
\advance\mins by \hours
\multiply\mins by 60
\multiply\hours by 100
\countdef\miltime=203
\advance\miltime by \hours
\advance\miltime by \time
\advance\miltime by -\mins

\begin{document}

\title{The Variability of Gamma-Ray Bursts that Create Afterglows}
        

\author{Joseph A. Mu\~{n}oz$^1$ and Jonathan C. Tan$^{1,2}$
        }
\affil{1. Princeton University Observatory, Peyton Hall, Princeton,
NJ 08544, USA\\ 
2. Inst. of Astronomy, Dept. of Physics, ETH Z\"urich, H\"onggerberg, 8093 Z\"urich, Switzerland\\
{\tt jmunoz@princeton.edu, jt@astro.princeton.edu}}

\begin{abstract}
  We consider whether the variability properties of Gamma-Ray Bursts
  (GRBs) that produce bright optical and longer wavelength transient
  afterglows (A-GRBs) are the same as a larger, inclusive sample of
  bright, long-duration GRBs, selected only by their $\gamma$-ray
  emission.  This sample may include a significant population of
  physically distinct ``dark'' or ``faint-afterglow'' GRBs with
  different variability properties, or may be composed of a single
  population, some of which lack afterglows only because of
  observational selection effects. We argue that the structure
  function is the most appropriate method for measuring the
  variability of bursts because of their transient and aperiodic
  nature. We define a simple statistic: the ratio of the integrated
  structure function from 0.1 to 1~s compared to that from 0.1 to
  10~s, as measured in the observer frame. To avoid instrumental
  effects we restrict our analysis to GRBs with BATSE data. Comparing
  10 A-GRBs to a ``main'' sample of about 500 bursts, we find there is
  a probability of only 0.03 of the samples being drawn from the same
  population, with the A-GRBs tending to have relatively less power on
  sub-second timescales. We conclude that this result is tentative
  evidence for variations in the properties of GRB progenitors that
  affect both the gamma-ray and afterglow properties of long-duration
  GRBs. In addition, our method of analyzing variability identifies a
  characteristic timescale of $\sim 1$~s, below which variability is
  suppressed, and finds a trend of increased short timescale
  variability at higher $\gamma$-ray energies. The long-duration GRBs
  that we identify as having the most sub-second timescale
  variability, may be relatively bright examples of short-duration GRBs.
\end{abstract}


\section{Introduction\label{sec:intro}}
      
The origin of Gamma-Ray Bursts (GRBs) continues to be a challenging
problem. Observationally there is now clear evidence linking {\it
  some} long-duration GRBs with the deaths of massive stars in
explosions known as supernovae or hypernovae.
GRB~980425 was likely associated with SN~1998bw (Galama et al. 1998),
while the connection of GRB~030329 with SN~2003dh is even more
convincing (Stanek et al.  2003).  There is also tentative evidence
associating GRBs that produce longer-wavelength transient
afterglows (A-GRBs) with galactic star-forming regions (Kulkarni et
al. 2000; Djorgovski et al. 2001; Bloom et al.  2002; Frail et al.
2002; Djorgovski et al. 2003).

The observed nonthermal $\gamma$-ray spectra of GRBs and their short
variability timescales necessitate relativistic motion of the material
producing the photons (Fenimore et al. 1993; Woods \& Loeb 1995; Tan,
Matzner \& McKee 2001; Lithwick \& Sari 2001).  The deceleration of
this ejecta by the surrounding medium is then thought to produce the
observed longer-wavelength afterglows.

Creation of relativistic ejecta from massive stars is a major
theoretical challenge because of the complicated physics involved in
supernova explosions and the uncertain initial conditions - i.e. the
stellar structure just before core collapse. Overcoming the
``baryon-loading problem'' in these scenarios is an acute problem.  If
the relativistic ejecta producing the GRB acquire their high velocity
close to the forming black hole, then this engine must operate for a
time long enough for the outflow to first pierce and clear a path
through the stellar envelope (e.g. Aloy et al. 2000; Proga et al.
2003; Zhang et al. 2003; Matzner 2003). This is many dynamical
timescales of the inner black hole accretion disk. Alternatively, the
ejecta that eventually produce $\gamma$-rays may be accelerated to
relativistic speeds as the supernova explosion breaks out of the star
(Colgate 1974; Matzner \& McKee 1999; Tan et al. 2001). This last
process inevitably occurs at some level in central, energetic
explosions of stars.

The spectral and temporal properties of GRBs and their afterglows
produced in these scenarios may be quite distinct. In the first model
(i.e. involving production of ultra-relativistic ejecta near the
forming black hole) $\gamma$-rays are likely to be produced
predominantly from internal shocks in the jet or outflow, and the
variability may reflect timescales associated with the inner accretion
disk around the black hole or the hydrodynamic instabilities
associated with the interaction of the outflow with the stellar
envelope. In the second model (i.e. where the ejecta only become
ultra-relativistic far from the central engine, beyond the radius of
the pre-supernova star) $\gamma$-ray emission probably comes from an
external shock and variability will be on longer timescales, perhaps
associated with inhomogeneities in the external medium (e.g. Dermer \&
Mitman 1999). Such models have been criticized because it is difficult
for them to efficiently produce GRBs with very short timescale
variability (e.g. Piran 1999). Also Ramirez-Ruiz \& Fenimore (2000)
pointed out that there was no evidence that individual pulses within
GRBs became broader during the course of a burst, as would be expected
in the simplest external shock models.  On the other hand, Tan et al.
(2001) noted the actual GRB lightcurve variability timescale for many
bursts, particularly those with identified afterglows, was much longer
($\sim$ several seconds) than the canonical value ($\sim$ tens of
milliseconds) often considered: in other words most of gamma-ray
energy is liberated in relatively long duration temporal features.

In both of the above scenarios, the afterglow emission comes from
external shocks, but the properties of this emission depend on the
Lorentz factor, magnetization and density of the ejecta, as well as
the density of the ambient medium (M\'esz\'aros \& Rees 1997). These
could have quite different typical values depending on the progenitor
model, perhaps leading to large variations in the ratio of gamma-ray
to afterglow luminosity. 

In addition to models involving supernovae/hypernovae, there are other
models of GRB progenitors, such as neutron star mergers, that may
contribute to the observed GRB population. Again, it would seem to be
a great coincidence if these models and those produced from the core
collapse of a massive star had indistinguishable variability and
afterglow properties.

Long GRBs do show quite a wide variety in the properties of their
longer-wavelength afterglows.  While most long GRBs ($\sim 90\%$)
produce X-ray afterglows, De~Pasquale et al. (2003) concluded that
only about half have ``bright'' optical afterglows; the remainder have
been dubbed ``dark'' GRBs. The distinction between ``bright'' and
``dark'' in this context is not clear-cut: a number of authors have
argued that ``dark'' bursts are on average at least 2~mag fainter in
the $R$~band (Ghisellini, Lazzati, \& Covino 2000; Reichart \& Yost
2001; Lazzati et al. 2002). This is consistent with results from the
HETE2 satellite that suggest afterglows are actually present following
most bursts, but have a wide range of intrinsic luminosities (Lamb et
al. 2004). In the radio, Frail et al. (2003) found that 25 out of 75
GRBs had {\it detected} afterglows.

The faintness of afterglows may have a number of explanations.  In the
optical it could be due to extinction by dust, absorption by the
Ly$\alpha$ forest if the bursts are at high redshift, or due to the
afterglow being intrinsically faint due to particular properties of
the relativistic ejecta and ambient density. De~Pasquale et al. (2003)
did not observe the expected soft X-ray absorption that would be
present if obscuration is important in most of their sample of dark
bursts. If there are ``radio-dark'' GRBs then this property must be intrinsic
to the burst.

The goal of this study is to quantify GRB variability and compare a
sample of A-GRBs, (i.e. GRBs with relatively bright afterglow emission
corresponding to being in the upper $\sim 1/2$ of their afterglow
luminosity functions), with a larger, inclusive ``main'' sample of
bursts, selected only because they are quite bright in gamma-rays and
of relatively long duration. To reduce the effects of observations
with different instruments we restrict our attention to GRBs with
BATSE data.

Given the state of theoretical modeling, it is unclear {\it a priori}
whether afterglow emission strength will correlate with particular
variability properties. However, as discussed above, the existence of
such a correlation would not be too surprising. If we find that A-GRBs
have statistically distinguishable variability properties compared to
the main sample, then this is evidence for intrinsic variations in the
progenitor model that affect both the gamma-ray emission and the
subsequent afterglow. The variation could take the form of a smooth
continuum of progenitor properties or of distinct sub-classes. If no
difference is seen then either there is only one type of progenitor,
or the different types do not lead to correlated variations in
afterglow variability and afterglow properties.

\begin{deluxetable}{cccccc} 
\tablecaption{Afterglow-GRBs observed by BATSE\label{tab:aglow}}
\tablewidth{0pt}
\tablehead{
\colhead{GRB no.} & \colhead{BATSE trig. no.} & \colhead{Redshift} & \colhead{$C_{\rm peak} ({\rm cts/64ms})$} & \colhead{$A_1/A_{\rm tot}$} & \colhead{$(A_1/A_{\rm tot})_0$}
}
\startdata
GRB 970508 & 6225 & 0.8349 & 172 & 0.0225 & 0.0285\\
GRB 970828 & 6350 & 0.9578 & 1640 & 0.0148 & 0.0155\\
GRB 971214 & 6533 & 3.42 & 404 & 0.0254 & n/a\\
GRB 980425 & 6707 & 0.0085 & 176 & 0.00843 & 0.00845\\
GRB 980703 & 6891 & 0.9662 & 396 & 0.0111 & 0.0107\\
GRB 990123 & 7343 & 1.6004 & 2990 & 0.00566 & 0.00972\\
GRB 990506 & 7549 & 1.3066 & 3490 & 0.0167 & 0.0243\\
GRB 990510 & 7560 & 1.6187 & 2010 & 0.0254 & 0.0357\\
GRB 991216 & 7906 & 1.02 & 13800 & 0.0224 & 0.0215\\
GRB 000131 & 7975 & 4.511 & 1620 & 0.0227 & 0.0355\\
\enddata
\end{deluxetable}

\newpage
\section{Measures of GRB Variability\label{sec:measures}}

We shall use the structure function to characterize GRB lightcurve
variability. First we review other statistical measures of variability
that have been employed in GRB studies. We then motivate our choice of
the structure function.


Beloborodov, Stern, \& Svensson (2000; hereafter BSS) studied the
power density spectra (PDS), $P_f=F_fF_f^*$ where $F_f$ is the Fourier
transform of the peak-normalized lightcurve, $F(t)$, of BATSE GRBs.
Their sample of about 500 bursts was defined by requiring that
$T_{90}$, the time to accumulate from 5\% to 95\% of the total fluence
in the 4 BATSE energy channels was greater than 20~s, that the peak
count rate in channels 2+3 (55-110 and 110-320~keV) was $>100$
counts per 0.064~s time bin, and that the total fluence was greater
than 32 times that of the peak time bin. They concluded the PDS of
bursts was well-described by a power law, $P_f\propto f^{-5/3}$ for
$f<1$~Hz. They also found evidence for a break in the power law at
higher frequencies, such that the PDS declined more rapidly. The PDS
method based on the Fourier transform has disadvantages because GRBs
are aperiodic and transient signals, with varying durations. To
adjust the PDS of each burst to a uniform frequency range, BSS added
artificial zero flux portions to the lightcurve so that the total
duration was 1048~s. As they comment, this ``zero padding'' introduces
an artificial, fluctuating contribution to the PDS that can dominate
the true signal at low frequencies.

Shen \& Song (2003) analyzed the variability of a similar sample of
bright, long-duration GRBs by calculating the variation power defined
as $P(\tau)=(1/N)\Sigma_{i=1}^{N} (m_i - \bar{m})^2/\tau^2$, where
$m_i, \:i=1,...,N$ is a counting series obtained from the time history
of the observed photons with a time step $\tau$, including background
photons. The power density is calculated via $p(\tau) = P(\tau_1) -
P(\tau_2)/(\tau_2-\tau_1)$. The power density was calculated for a
noise series for each burst and then subtracted from $p(\tau)$. Shen
\& Song found the timescale at which $p$ was a maximum for each burst
and claimed evidence for a bimodal distribution in these peak
timescales, with roughly half the bursts peaking at $\tau<1$~s and
half peaking at $\tau>1$~s. We performed this method of analysis on a
similar sample of BATSE GRBs (our main sample), finding there were a
significant number of bursts where the peak of $p(\tau)$ was at the
minimum timescale set by the BATSE time resolution of 64~ms. Also many
bursts had a relatively broad and flat-peaked power density, which
is not reflected in the single value of the timescale of the peak. For
these reasons we have preferred to use our structure function
analysis, described below.

Borgonovo (2004) studied the discrete autocorrelation function (ACF,
the Fourier transform of the PDS), defined as $A(\tau = k\Delta T) =
\Sigma_{i=0}^{N-1} c_ic_{i+k}/A_0,\:k=1,...,N-1$, where $c_i$ is the
number of counts in a given time bin after subtraction of background
$b_i$ and $A_0\equiv \Sigma_{i=0}^{N-1} c_i^2 - (c_i+b_i)$, of a
sample of GRBs with optical afterglows and known redshifts, focusing
mainly on those with BATSE data (as in our A-GRB sample). He
considered the half width at half maximum of the ACF (corrected for
cosmic redshift), claiming evidence for a bimodality in the
distribution. However, differences in this width of the ACF depend
quite sensitively on the finite duration of the bursts, which are
influenced by observational selection effects, and less sensitively on
the actual variability properties at a fixed timescale.

Reichart et al. (2001) studied the variability of bursts with known
redshifts by looking at the (mean summed squared) difference between
the observed lightcurve and the same lightcurve smoothed on a
particular timescale, proportional to the burst duration. Using 11
bursts with measured variabilities and isotropic peak luminosities,
the value of the smoothing timescale was optimized to give the largest
change in isotropic peak luminosity for a given change in variability.
By this method a significant correlation between this measure of
variability and luminosity was found. One important feature of this
method is the use of smoothing timescales that are proportional to
burst duration rather than a fixed timescale. Reichart et al.  (2001)
noted that if fixed timescales were used then no correlation was
found. 

GRB lightcurves have also been studied by decomposition into pulses of
a particular functional form (e.g. Norris et al. 1996; Ramirez-Ruiz \&
Fenimore 2000; Lee, Bloom \& Petrosian 2000; Nakar \& Piran 2002;
Quilligan et al. 2003).  This method suffers because it requires an
arbitrary specification of the functional form of the pulses, but has
the advantage that once this assumption has been made, relatively good
statistics on pulse properties can be obtained from a given set of
burst data.  Ramirez-Ruiz \& Fenimore (2000) looked at pulse width
evolution within GRBs finding that pulses did not appear to get
broader over the course of bursts, contrary to the simplest models of
production in external shocks. Lee et al.  (2000) found that pulse
timescales tend to be shorter in GRBs with higher peak fluxes, as
expected from cosmic time dilation, although there is also tentative
evidence for a contribution from processes intrinsic to the bursts.






\subsection{The Structure Function Analysis}

We use the first-order structure function (Rutman 1978), defined as
\beq
\label{sfn}
D^1(\tau) \equiv \langle \left[ F(t) - F(t+\tau)\right]^2 \rangle,
\eeq where the angular brackets denote an ensemble average, $F(t)$ is
the flux at time $t$ and $\tau$ is the time lag. For a stationary
random process $D^1(\tau) = 2\sigma^2[1-\rho(\tau)]$, where $\sigma^2$
is the variance and $\rho(\tau)$ is the autocorrelation function.
Although this is only approximately valid for GRBs, to first order we
expect that at short $\tau$, $D^1$ tends to a constant value that is
twice the variance of the measurement noise (i.e. if there were no
noise it would tend to zero). For $\tau$ longer than the longest
correlation time scale, $D^1$ tends to a value equal to twice the
variance of the fluctuation. However, the burst nature of GRBs means
that the longest correlation time scale is half the burst duration.
For example, for a burst that is 64~s long, the largest value of
$\tau$ considered is 32~s. As the time bin is 64~ms, there are 1000
flux measurements and $D^1(\tau=32\:{\rm s})$ is based on an average
of 500 correlations. $D^1(\tau=64\:{\rm ms})$ is based on an average
of 999 correlations.

We take the sample of GRBs selected by BSS as the basis of our
main sample. We do not impose the fluence condition (see above)
that BSS implemented that affects about 5\% of their sample.  We do
exclude some GRBs lacking good background estimates or full coverage
at 64~ms resolution.
Our main sample contains
450 bursts. Note that the BSS sample does not include any bursts
detected after 17th June 1998. The A-GRB sample (see Table \ref{tab:aglow}) is made up of 10 bursts with
BATSE data and for which afterglow counterparts have been detected.
The ability to detect afterglows was developed relatively late in the
BATSE mission. Three of the bursts (with trigger nos. 6225, 6533 and
6707) are in both samples, but all of the A-GRBs meet the
flux requirements of the main sample.


To define GRB lightcurves from the BATSE data we sum the counts in
channels 2 and 3, corresponding to an energy range of
$\sim$55-320~keV, as in the study of BSS.  We subtract a background,
utilizing published background
fits\footnote{http://cossc.gsfc.nasa.gov/batseburst/sixtyfour\_ms/bckgnd\_fits.html}
and then normalize the lightcurves so that the amplitude at peak is
unity (BSS). The start of each burst is defined by the BATSE trigger
time.  We define the end of the burst by working backwards from the
end of the available data stream to the point when the flux (measured
in a single bin) is 5\% of its peak value and when the absolute flux
(average over 1~s) is $\geq 300\:{\rm cts\:s^{-1}}$. The latter
condition was imposed to avoid the effects of Poisson fluctuations in
noisy bursts. We then set the end of the burst so that the length of
the burst is 20\% longer than this interval. If the burst duration is
still less than 50~s, then we extend the end of the burst to 50~s
after the trigger time. Note that for inclusion in our sample the
bursts must have $T_{90}>20$~s, evaluated over all energy channels
(BSS). We evaluate $D^1(\tau)$ for each burst. As the time resolution
of the data is 0.064~s, the range of $\tau$ is thus 0.064~s to at
least 25~s.

\begin{figure}[ht]
\epsfig{file=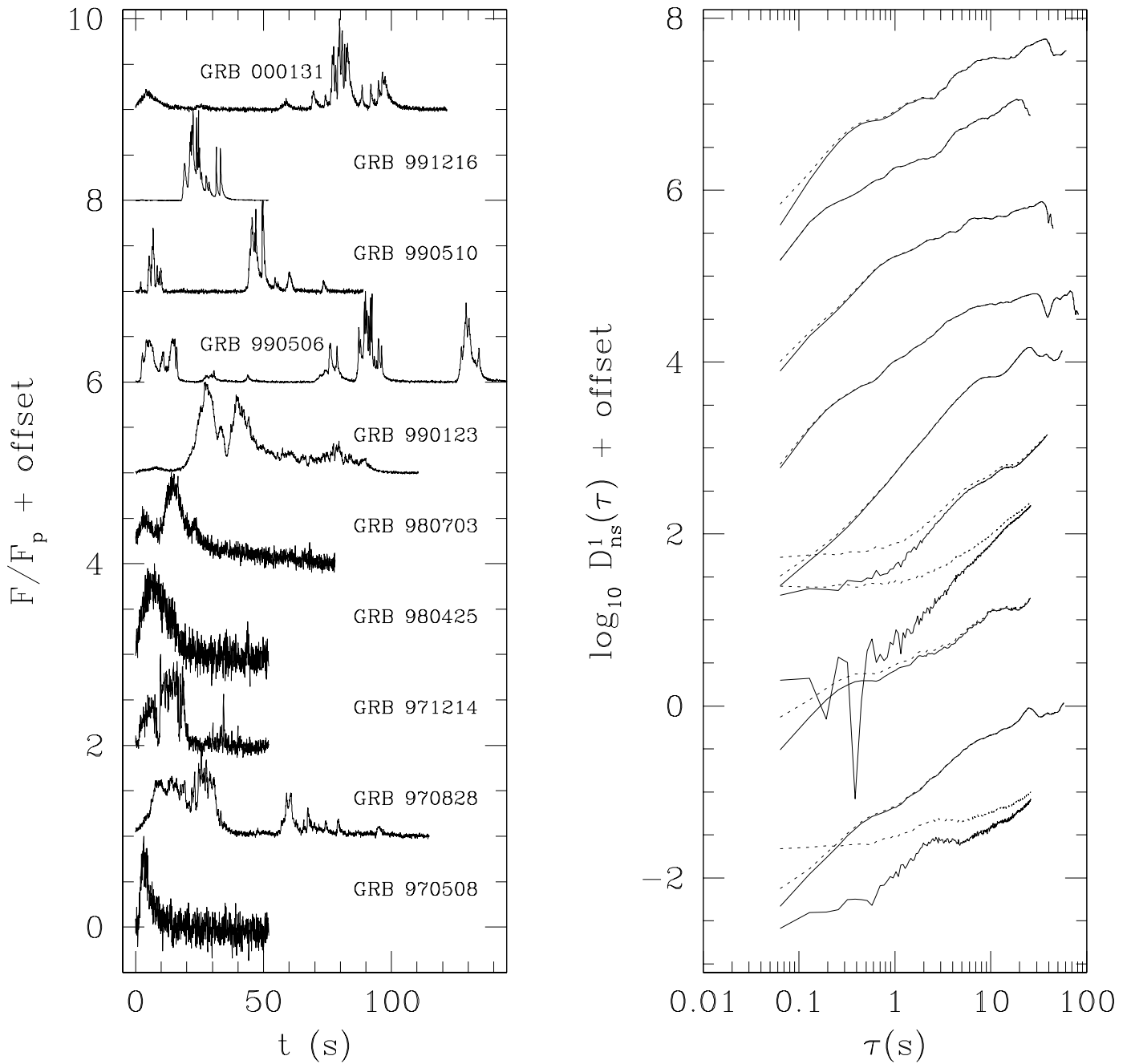,height=7.5in,width=7.5in}
\figcaption{\label{fig:lcminaglow} Left panel: BATSE light curves
  (peak-normalized, channels 2+3, offsets $0,+1,...,+9$) of the A-GRB
  sample (the final $\sim 20$~s of GRB~990506 are not shown). (b)
  Right panel: Noise-subtracted structure functions, $D^1_{\rm
    ns}({\tau})$ (solid lines) of the same bursts (offsets
  $0,+1,...,+9$). We show the structure functions before noise
  subtraction with dotted lines.  }
\end{figure}

We then define a sample of noise of the same duration and
normalization for each burst by using the data following the above
interval. We take the structure function of the noise and subtract
this from the structure function of the burst to derive the
``noise-subtracted'' structure function, $D^1_{\rm ns}$. This
procedure is illustrated in Figure~\ref{fig:lcminaglow}.  The
noise-subtracted structure function does not tend to exhibit the
plateau at small $\tau$. Discreteness associated with the 64~ms time
resolution of BATSE is evident.

\begin{figure}[ht]
\begin{center}
\epsfig{file=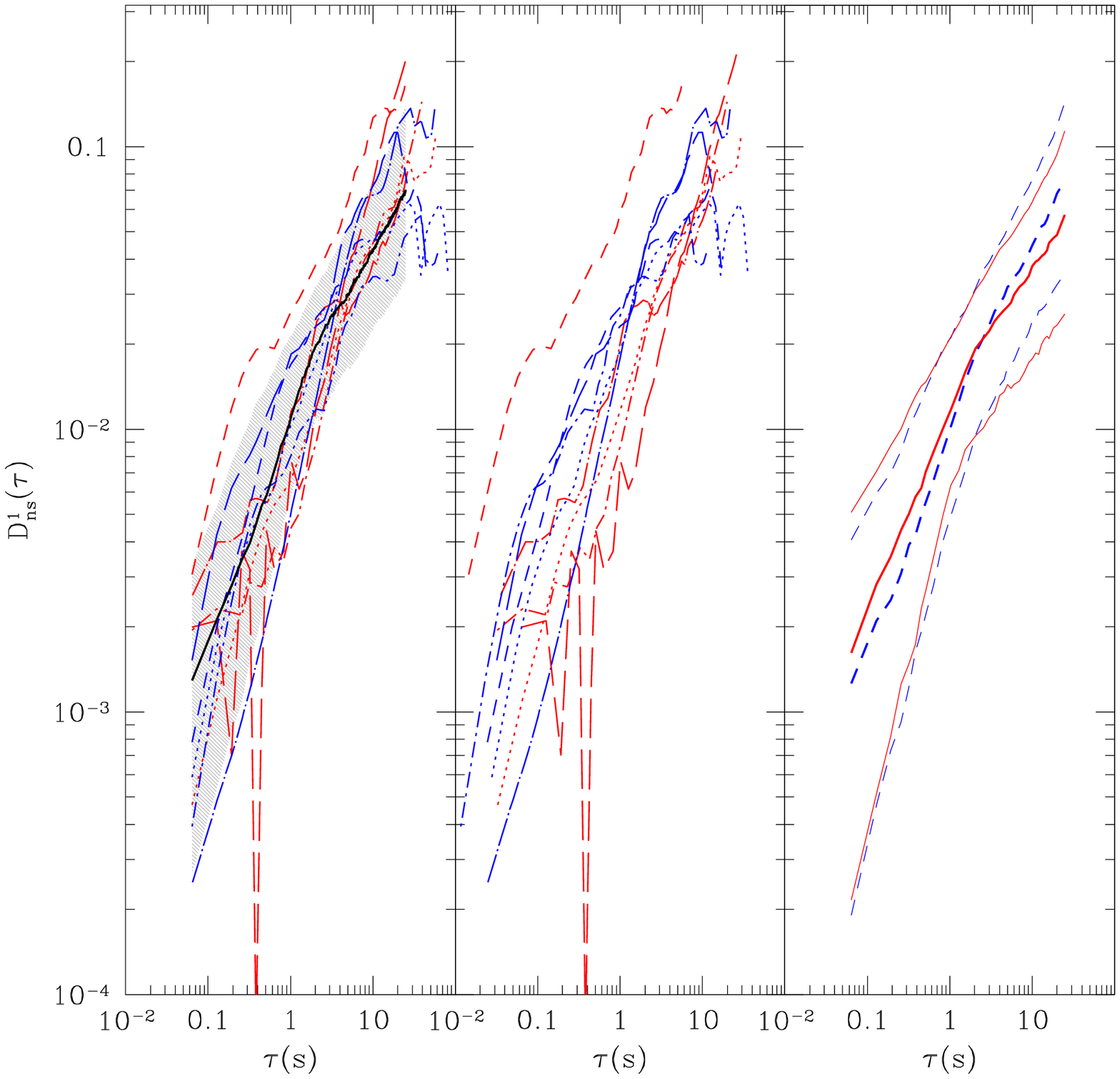,height=4in,width=7.5in}
\end{center}
\caption{\label{fig:aglow} (a) Left panel: Noise-subtracted
  structure functions, $D^1_{\rm ns}({\tau})$, for the A-GRB sample
  (various non-solid line types).  The median of the main sample is
  shown by the heavy, solid line, and the interval containing 68\% of
  this sample is shown by the shaded region. (b) Middle panel: the
  same as (a), but now only showing the A-GRBs corrected for
  redshifting of the variability timescales. (c) Right panel: the same
  as (a), but now comparing the main sample at lower (25-110~keV;
  channels 1+2 --- dashed heavy line is the median and the upper and
  lower thin dashed lines delineate 68\% of the sample) and higher
  ($>$110~keV; channels 3+4 --- equivalent red solid lines) observed
  energies. At higher energies the GRBs have relatively more
  sub-second scale variability.  }
\end{figure}

In Figure~\ref{fig:aglow}a we show the noise-subtracted structure
functions for the 10 A-GRBs together with the median and 68~percentile
bounds (evaluated at each value of $\tau$) of the distribution of same
functions of the main sample. 

The slope of the median of the structure functions of the main sample
changes fairly abruptly at $\tau \simeq 1-2$~s.  This is a similar
timescale to the break in the power density spectra reported by BSS.
As these authors note, if this signal is produced in the rest frame of
a relativistic outflow, then one would expect variations of the
outflow Lorentz factor, $\Gamma$, to smear out the break.  They
estimate that the dispersion in $\Gamma$ would need to be
$\Delta\Gamma/\Gamma\lesssim 2$ to preserve the observed break, but
regard such a narrow dispersion as being unlikely.  Alternatively, BSS
comment that the break timescale could be associated with the central
engine.  However, $1-2$~s is orders of magnitude longer than the
dynamical timescale of the central engine (presuming it is a stellar
mass black hole). Finally BSS comment that the break timescale may be
associated with the observed dynamical timescale of the region where
the outflow becomes optically thin, i.e. at the photospheric radius,
$R_p$. Variability on timescales shorter than $t_p = R_p /
(c\Gamma^2)$ is suppressed. Again, it appears that a narrow range of
$\Gamma$ would be needed to preserve the sharpness of the break.  The
same issue applies to a model of generating variability from
interaction of the outflow with density inhomogeneities in a wind
beyond $R_p$. Further study, preferably with larger samples, is needed
to confirm whether there is indeed a characteristic observer-frame
variability timescale of $1-2$~s. Such a timescale would be a
challenging constraint on progenitor models.


The redshifts of the A-GRBs are known so we can correct for
cosmological time dilation (Fig.~\ref{fig:aglow}b).  Note, however,
that this analysis does not account for the use of a fixed observed
energy range, $\sim$55-320~keV, to define the lightcurves. The
variability properties of lightcurves are likely to depend on the rest
frame energy of the emission.  To examine this issue, we considered
the structure functions of main sample burst light curves defined with
observed energies 25-110~keV (channels 1+2) and $>$110~keV (channels
3+4) (Fig.~\ref{fig:aglow}c). We find that bursts have relatively more
short (sub-second) timescale variability at higher energies than at
lower.  This is consistent with the results of Fenimore et al. (1995),
who studied the dependence of autocorrelation function width with
observed energy for 45 bright GRBs observed by BATSE and found it to
be $\propto E^{-0.4}$.



In order to characterize the relative amounts of power on short and
long timescales we consider the area under the structure function for
$0.1<\tau<10$~s. All the bursts in our samples have a well-defined
value of $D^1_{\rm ns}(\tau)$ over this regime. As a very simple
measure of relative variability, we consider the ratio of $A_1\equiv
\int_{0.1\:{\rm s}}^{1\:{\rm s}} D^1_{\rm ns}(\tau) \: d\tau$ to
$A_{\rm tot} \equiv \int_{0.1\:{\rm s}}^{10\:{\rm s}} D^1_{\rm
  ns}(\tau) \: d\tau$.  For most of the afterglow sample we can also
determine the corresponding area ratio for $D^1_{\rm ns}(\tau_0)$
(this is not possible only for GRB~971214, a relatively short burst at
high redshift). The uncertainties in $A_1/A_{\rm tot}$ are largest for
noisy bursts that have a relatively small value of $A_1$, the value of
which is quite strongly influenced by the noise subtraction.

\begin{figure}[ht]
  \centerline{\epsfig{file=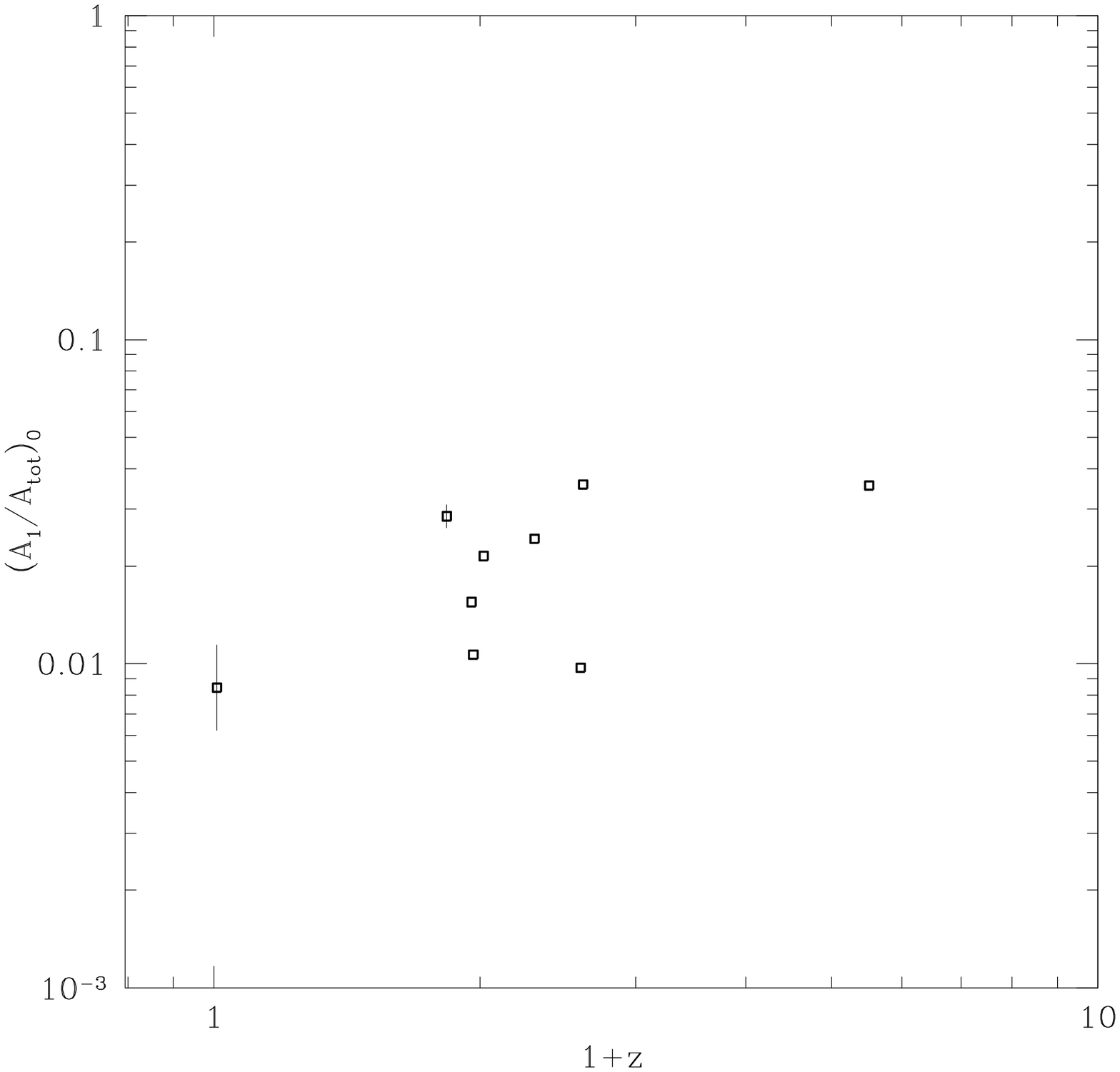,width=\figwidth}}
  \figcaption{\label{fig:zcorel} The dependence of rest frame area
    ratio, $(A_1/A_{\rm tot})_0$, of structure functions of A-GRBs
    with $1+z$, or equivalently the rest frame energy that is probed
    by the BATSE observations. For those bursts with sufficient
    post-burst data to create more than one noise sample, the error
    bars indicate the uncertainty introduced by the noise subtraction. These are only significant for GRB~980425 and GRB~970508.
}
\end{figure}

Given the trends in variability with observed energy, as determined
from the large main sample (Fig.~\ref{fig:aglow}c), we expect that
higher-redshift bursts in the A-GRB sample, which are being
observed at higher rest-frame energies, should exhibit shorter
variability timescales and have relatively more power on shorter
timescales. We show the dependence of the rest frame area ratio with
$1+z$ in Fig.~\ref{fig:zcorel}. There is an apparent weak correlation of
increased sub-second variability with $1+z$, but this effect is
dominated by the lowest and highest redshift bursts in the sample. 
At any redshift, we expect quite a broad distribution in $(A_1/A_{\rm
  tot})_0$, so a larger sample is required before one can look for a
statistically significant correlation, particularly as GRB~980425 may
be very different in nature from the other bursts (e.g. because of its
small $\gamma$-ray luminosity, Galama et al. 1998).

\begin{figure}[ht]
\centerline{\epsfig{file=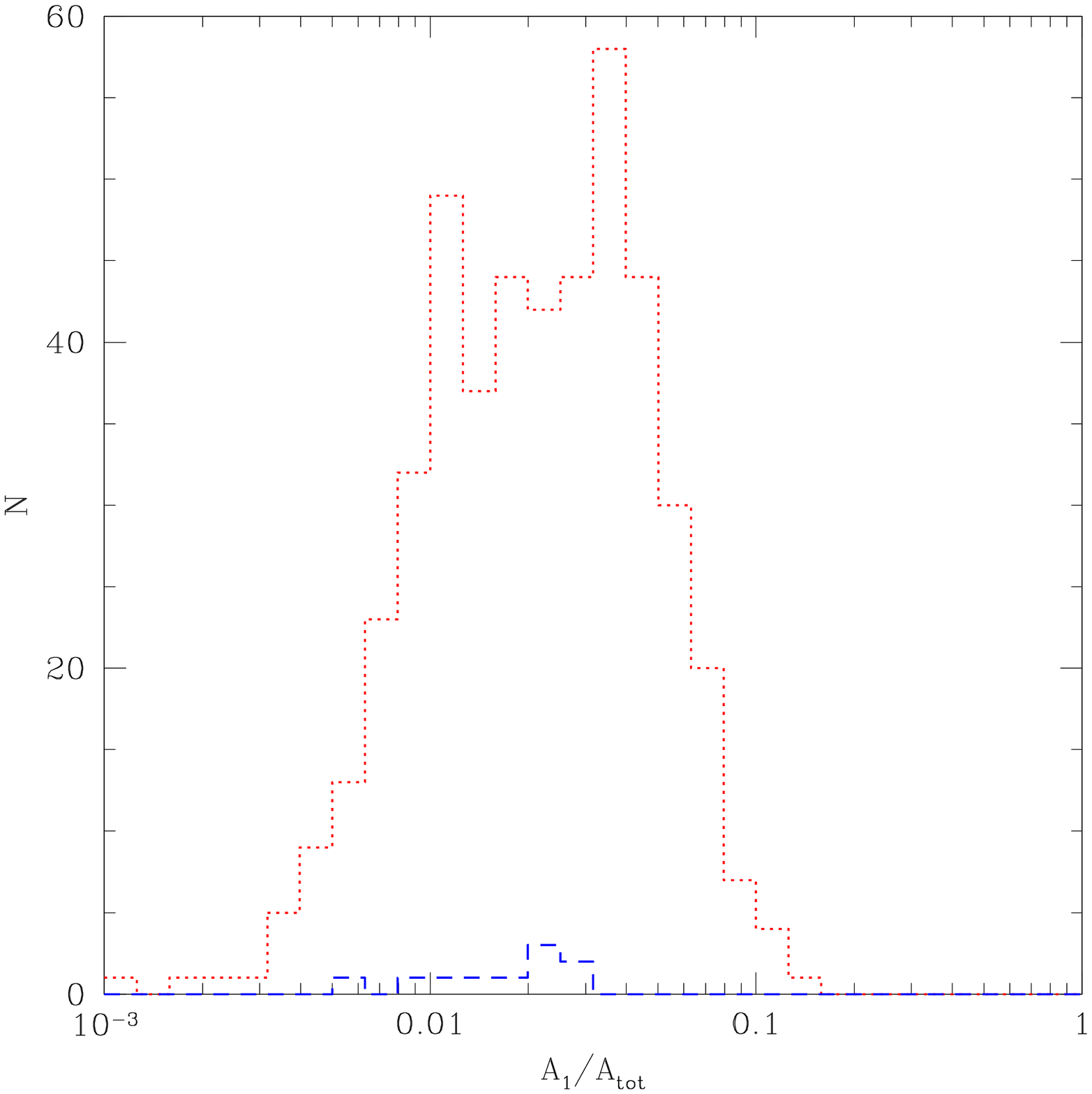,width=\figwidth}}
\figcaption{\label{fig:sfhist}
Histograms of area ratio $A_1/A_{\rm tot}$ of structure functions from A-GRB (dashed) and main (dotted) samples. 
}
\end{figure}

The distributions of $A_1/A_{\rm tot}$ are shown in Figure
\ref{fig:sfhist} for the A-GRB and main samples (see also Table
\ref{tab:aglow} for the A-GRBs).  We performed a Kolmogorov-Smirnov
(KS) test on these two data sets.  This yielded a probability of 0.030
that they are drawn from the same distribution. The sense of the
discrepancy can be seen from Figure~\ref{fig:sfhist}: there are no
A-GRBs with relatively high values of $A_1/A_{\rm tot}$, i.e. they are
relatively lacking in sub-second scale power. This result agrees with
the qualitative assertion by Tan et al. (2001) that ``most of the GRBs
with afterglows have relatively smooth pulses''. 

The formal probability of 0.030 that the A-GRB and main samples can be
drawn from the same distribution is small, but not negligible.
Several factors make it relatively difficult to pick out a distinct
population of ``dark'' GRBs, i.e. GRBs with faint afterglows, even if
one is present in the main sample: the main and A-GRB distributions of
$A_1/A_{\rm tot}$ are quite broad; the samples overlap, so that the
main sample probably contains a significant fraction of bursts that do
create optical afterglows (this is also implied by the results of
De~Pasquale et al. 2003 and Lamb et al. 2004); the A-GRB sample is
small; there may be differences introduced if the A-GRB and main
samples have significantly different redshift distributions (we are
forced to assume that these are the same so that the effects of cosmic
time dilation and the dependence of variability on rest frame energy
are averaged out).  The test could obviously be improved if the
samples (particularly of A-GRBs) were larger. Although about 20
additional A-GRBs have been observed with instruments other than
BATSE, many of these bursts have lower signal-to-noise because of the
smaller effective areas of the other instruments. Additionally, not
all the data are publicly available at present. In restricting our
analysis to only the BATSE sample we also circumvent problems due to
different instrumental selection effects between the A-GRB and main
samples.

Is the apparent discrepancy between the A-GRB and main samples an
artifact of the particular method of comparison? The area ratio
$A_1/A_{\rm tot}$ is a particularly simple way of measuring relative
variability over these timescales. We have also repeated the analysis
for ratios of the areas defined from 0.2 to 2~s and 0.2 to 20~s. The
probability of a single parent distribution remains small at 0.029,
but this is obviously not a completely independent test.  Although the
structure functions are not particularly well approximated by single
power laws, we have found the best fitting indices for each burst
(weighting the structure function uniformly in $\rm log\:\tau$) and
compared these distributions. We find that they are not significantly
different. This illustrates the difficulty of trying to characterize
the variability properties of bursts with a single number.

\begin{figure}[ht]
\epsfig{file=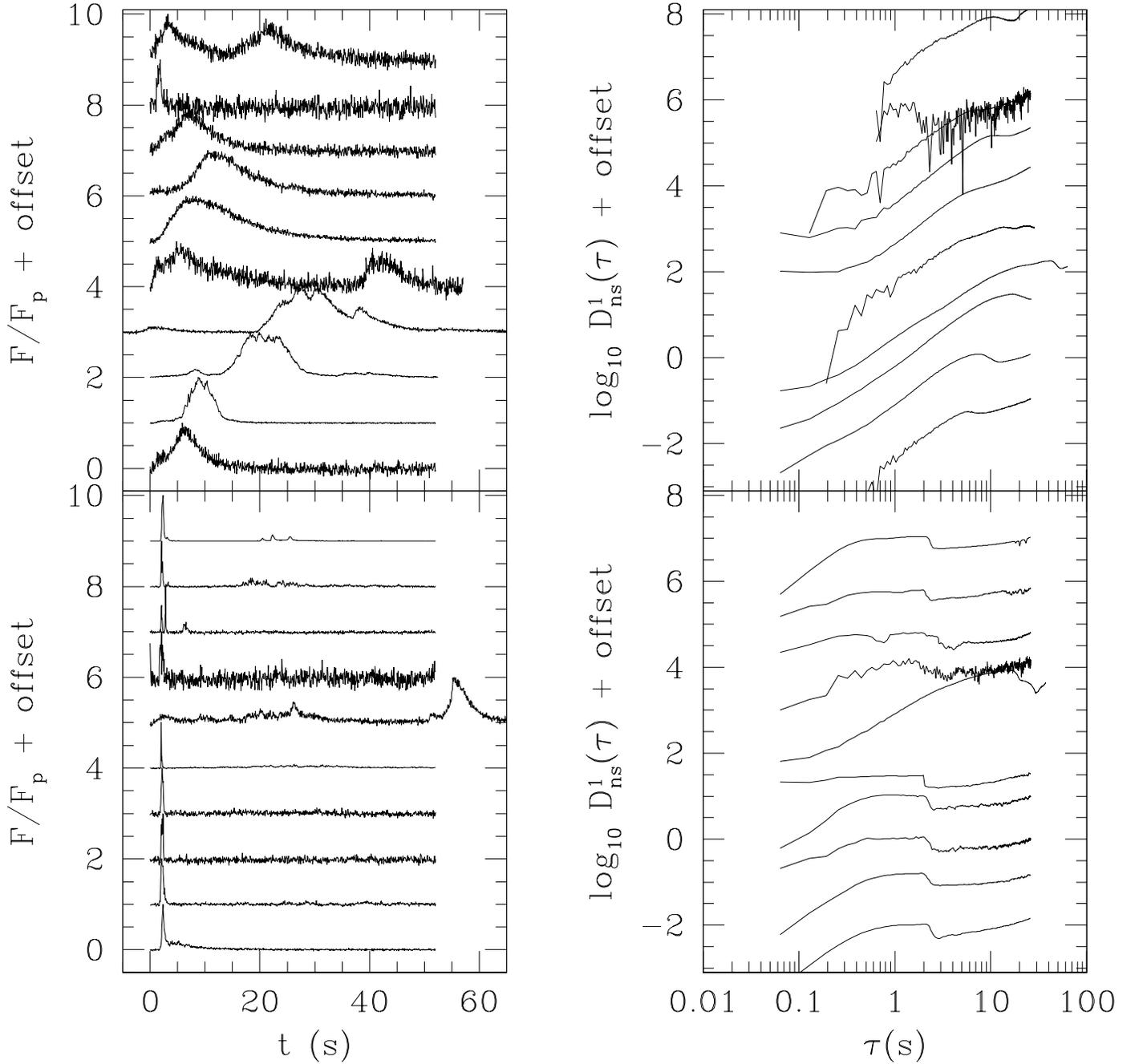,height=7.5in,width=7.5in}
\figcaption{\label{fig:lcmin}
Peak-normalized lightcurves (left panels, with offsets $0,+1,...,+9$) of samples of GRBs with the minimum (top panel: BATSE trigger nos. 1039, 1886, 2067, 2138 (shifted also by $+50$~s so first weak pulse is not shown), 2276, 2387, 3003, 5531, 5723, 6167) and maximum (bottom panel: BATSE trigger nos. 0467, 0503, 1546, 1626, 1997, 2700, 3248, 5572, 5725, 5989) area ratios $A_1/A_{\rm tot}$ of their structure functions (displayed in the right panels with offsets of $0,+1,...,+9$).
}
\end{figure}

In order to gain more insight into the nature of variability
differences that are indicated by the $A_1/A_{\rm tot}$ statistic, in
Figure~\ref{fig:lcmin} we show the lightcurves and structure functions
of the ten bursts with the lowest and highest values of $A_1/A_{\rm
  tot}$ from the main sample. The bursts with the smallest value of
the area ratio (i.e. those with relatively less 0.1~s to 1.0~s scale
variability) indeed appear to be very ``smooth'' (remember that
variation due to noise fluctuations has been subtracted off the
structure functions). There is only one case out of ten (Trigger no.
5723) where this does not appear to be the case and the burst is in
fact dominated by quite short timescale features. The low area ratio
of this burst is probably due to an uncertain and large correction for
the variability due to noise. 

Now consider bursts with the largest values of the area ratio: nine
out of ten exhibit a single, very strong, short-timescale initial
pulse (which sets the peak normalization), followed by relatively
minor emission features. This shows that the $A_1/A_{\rm tot}$
statistic is working to separate GRBs with different relative amounts
of sub-second timescale variability.  Also since the morphologies of
the bursts with the largest values of $A_1/A_{\rm tot}$ are quite
homogeneous as a group, forming quite a distinct subset of the range
of morphologies exhibited by long-duration GRBs, it is possible that
this is caused by real physical differences in their progenitors. Such
differences may also explain why none of the A-GRBs, i.e.
``bright-afterglow'' GRBs, populate this region of $A_1/A_{\rm tot}$
parameter space or exhibit this morphology.  We note that some of
these bursts, if observed with lower signal to noise, might be
classified as short-duration bursts, i.e.  with $T_{90}\lesssim 2$~s.
This interpretation is consistent with the finding that short-duration
GRBs do in fact have faint, longer-duration hard x-ray tails (Lazzati,
Ramirez-Ruiz, \& Ghisellini 2001), as seen by co-adding a large number
of short bursts. Another partial explanation for the lack of detected
afterglows from these types of bursts may then be instrumental, as the
localizations of short bursts by previous and current satellites have
been relatively poor compared to longer bursts.

What are the lightcurve and variability properties of the ``dark'' GRBs,
as identified by De~Pasquale et al. (2003)? First note that many of
these bursts may be optically dark in their afterglows because they
are faint in all their emission properties: De~Pasquale et al. find
the X-ray fluxes of A-GRBs are about 5 times higher than the dark
bursts --- if the optical-to-X-ray fluxes are constant then 75\% of
the dark bursts have predicted optical fluxes below the level to which
they were observed.  Conversely, about 25\% of the dark bursts have
optical-to-X-ray flux ratios that are at least 4-10 times smaller than
the A-GRBs. A lack of soft X-ray absorption suggests that this
faintness is not due to obscuration.  The $\gamma$-ray and hard X-ray
light curves of the dark bursts listed by De~Pasquale et al. (2003)
are very diverse, including both short ($\sim$~few seconds) and long
($\sim$~hundreds of seconds) bursts. Of the 20 dark bursts, four were
observed by BATSE: GRBs 970111, 971227, 990907, 991014 (with BATSE
trigger nos. 5773, 6546, 7755, 7803). The data for GRB~990907 are not
available in the current BATSE catalogue.  GRB~970111 is a relatively
smooth burst with multiple peaks and a 50~s duration, while the
remaining two bursts are quite short ($\sim 5-10$~s) with peaks with
rise and fall timescales $\lesssim 1$~s.  Thus some of the lightcurves
and variability properties of the dark bursts are similar to the
sample shown in the lower panel of Figure~\ref{fig:lcmin}, but some
are not, as is to be expected given the different reasons why
GRBs may be classified as optically dark.


\section{Conclusions}

After reviewing methods of assessing GRB variability, we have
presented tentative evidence that the variability properties, as
measured by the structure function, of A-GRBs, i.e. bursts with
detected afterglows, differ from the more general population of
long-duration GRBs. The A-GRBs have relatively little sub-second
variability.  This may be because the wider sample of GRBs contains a
significant number of ``dark'' bursts, i.e. with relatively faint
afterglows, that have a different physical origin and relatively more
shorter timescale ($<1$~s) variability.  It is also possible that our
result is explained by a more continuous trend from a single
progenitor model causing GRBs with brighter afterglows to have less
short timescale variability.

If the dark bursts were simply at higher redshifts then the above
effect might be due to the fact that we observe the bursts at a higher
rest frame energy, where variability timescales are shorter
(Fig.~\ref{fig:aglow}c).  However, cosmic time dilation would tend to
counteract this effect.  Furthermore, the example lightcurve
morphologies shown in Figure~\ref{fig:lcmin} are suggestive that there
are different physical mechanisms involved as the variability, as
measured by the $A_1/A_{\rm tot}$ statistic, changes.

To further test the above hypothesis, we need more stringent
constraints on the afterglow to $\gamma$-ray flux ratios of the bursts
with relatively high short timescale variability, such as those shown
in the lower panel of Figure~\ref{fig:lcmin}. This should be possible
with the {\it Swift} satellite.


In \S\ref{sec:intro} we argued that there are good physical reasons to
expect different classes of GRBs, depending on how long the forming
black hole can power relativistic jets and how long it takes these
jets to burrow through the stellar envelope. There are of course many
other theoretical models that may produce some of the observed GRBs,
such as the mergers of neutron stars. More work is needed on the
expected variability and afterglow properties of these models.

\acknowledgements We thank B. Draine, C. Matzner, C, McKee and E.
Rossi for helpful discussions, A. Beloborodov for providing a list of
the BSS sample of GRBs and Y. Kaneko for helping with the BATSE data
extraction for GRBs 970828 and 000131. We also thank the referee for
comments that greatly improved the paper. This work was supported in
part by NASA grant NAG5-10811. JCT has received support via a
Spitzer-Cotsen Fellowship from the Department of Astrophysical
Sciences and the Society of Fellows in the Liberal Arts of Princeton
University, and via a Zwicky Fellowship from the Inst. of Astronomy,
ETH Z\"urich.




\end{document}